\documentclass[onecolumn,showpacs,preprintnumbers,amsmath,amssymb]{revtex4}

\usepackage{tabularx}
\usepackage{array}
\usepackage{mathrsfs}
\usepackage{amssymb}
\usepackage{amsmath}
\usepackage{graphicx}
\usepackage{bm}
\usepackage{multirow}
\usepackage{float}
\usepackage{appendix}

\begin{document}
\title{Twin-field quantum digital signatures}
\author{Chun-Hui Zhang$^{1,2,3}$}
\author{Yu-Teng Fan$^{1,2,3}$}
\author{Chun-Mei Zhang$^{1,2,3,4}$}
\author{Guang-Can Guo$^{1,2,3,4}$}
\author{Qin Wang$^{1,2,3,4}$}\email{qinw@njupt.edu.cn}
\address{ $^1$ Institute of quantum information and technology, Nanjing University of Posts and Telecommunications, Nanjing 210003, China \\
 $^2$ Broadband Wireless Communication and Sensor Network Technology, Key Lab of Ministry of Education, NUPT, Nanjing 210003, China \\
 $^3$ Telecommunication and Networks, National Engineering Research Center, NUPT, Nanjing 210003, China\\
 $^4$ Key Laboratory of Quantum Information, CAS, University of Science and Technology of China, Hefei, Anhui 230026, China\\}

\date{\today}

\begin{abstract}
Digital signature is a key technique in information security, especially for identity authentications. Compared with classical correspondence, quantum digital signatures (QDSs) provide a considerably higher level of security, i.e., information-theoretic security. At present, its performance is limited by key generation protocols (e.g., BB84 or measurement-device-independent protocols), which are fundamentally limited in terms of channel capacity. Fortunately, the recently proposed twin-field quantum key distribution can overcome this limit. This paper presents a twin-field QDS protocol and details a corresponding security analysis. In its distribution stage, a specific key generation protocol, the sending-or-not-sending twin-field protocol, has been adopted and full parameter optimization method has been implemented.  Numerical simulation results show that the new protocol exhibits outstanding security and practicality compared with all other existing protocols. Therefore, the new protocol paves the way toward real-world applications of QDSs.
\end{abstract}

\pacs{03.67.Dd, 03.67.Hk, 42.65.Lm}

\maketitle

\section{Introduction}
Digital signatures (DSs) possess wide applications in validating the authenticity and integrity of digital documents such as financial transactions and electronic contracts. Present digital signatures, hereafter called classical digital signatures, possess security levels based on computational complexity. For example, the Rivest-Shamir-Adleman protocol \cite{RSA} relies on solving large number factorization problems, and the elliptic-curve-based protocol \cite{ECDSA1,ECDSA2} depends on discrete logarithms. Unfortunately, these protocols would all be cracked through the advancement of mathematical algorithms or the emergence of quantum computers.

In contrast, the security of quantum digital signatures (QDSs) is based on quantum mechanics laws. QDSs have proven to provide information-theoretic security and thus attracted a lot of attention from the scientific world. Since the first QDS protocol was proposed in 2001 \cite{Gottesman} and the first experimental demonstration was accomplished in 2012 \cite{Clarke}, many obstacles to practical applications have been removed (e.g., the demanding of quantum memories; releasing secure quantum channels \cite{Dunjko,Collins2014,Amiri}), such that the quantum key distribution protocols \cite{BB84} can be implemented in the key distribution stage \cite{Wallden}). Furthermore, measurement-device-independent (MDI) protocols \cite{MDIQDS,Roberts} were proposed to solve side-channel attacks on measurements. The passive protocol \cite{Qin1} can avoid information leakage that occurs during the intensity modulating process. To date, many theoretical and experimental studies have been examined this subject matter \cite{Donaldson,Collins2016,Croal,Yin2017,AnXB,Thornton}.

With present QDS protocols, a good balance between security and practical performance is still difficult to achieve. For example, one can obtain a higher signature rate with a lower level of security when using BB84-type QDS protocols. In contrast, MDI-type QDS protocols feature a higher level of security but worse signature rates. Most importantly, both protocol types cannot exceed the fundamental limit of channel capacities without quantum repeaters, as in QKD \cite{TGW,PLOB}. To overcome this fundamental limit in QKD, Lucamarini \emph{et al.} proposed the so-called twin-field quantum key distribution (TF-QKD) protocol \cite{TFQKD} and obtained excellent security and practical performance. Inspired by Lucamarini \emph{et al.}, we for the first time present a twin-field quantum digital signature (TF-QDS) protocol. We first implement a sending-or-not-sending (SNS) protocol \cite{SNSTF1,SNSTF2} into the key distribution stage, and analyze its security against general attacks within QDSs (e.g., forging and repudiation attacks) by taking finite-size-effects into account. Furthermore, we perform corresponding numerical simulations and full parameter optimizations. When compared to other protocols such as BB84-QDS \cite{Amiri} and MDI-QDS \cite{MDIQDS}, our method exhibits outstanding results in terms of signature rates and transmission distances.

\section{Twin-field QDS}\label{sec2}
\subsection{Protocol procedure}\label{protocol}
\begin{figure*}[htbp]
\centering
\includegraphics[width=12cm]{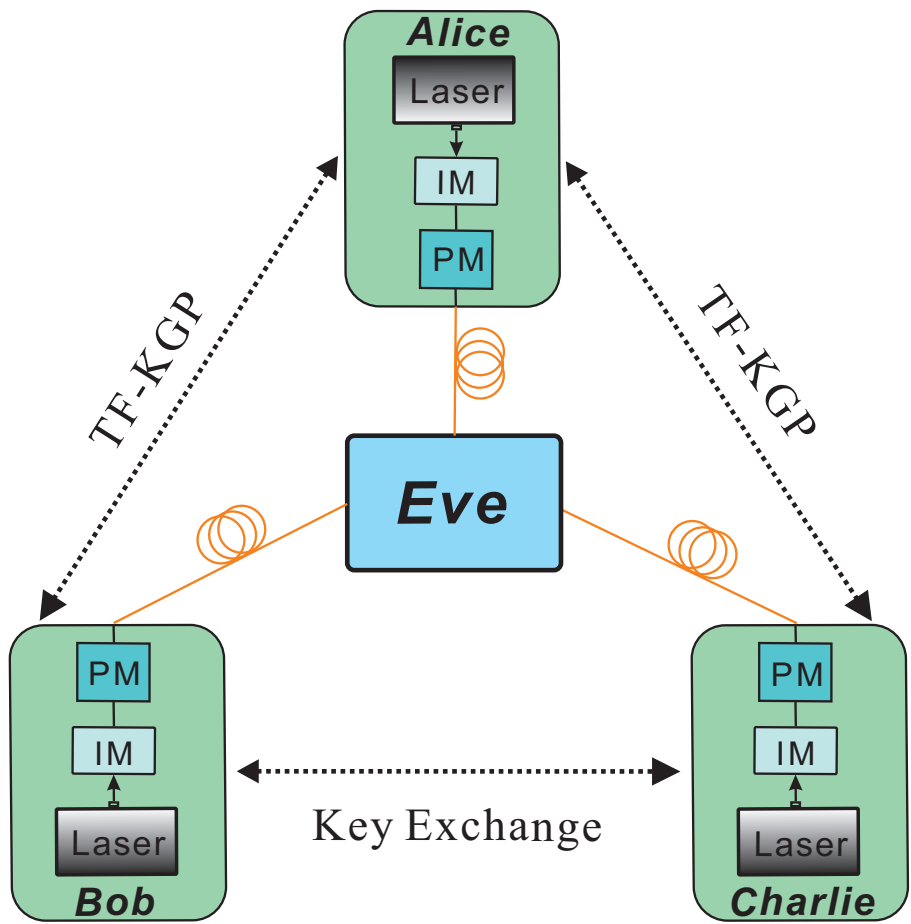}
\caption{Schematic of the TF-QDS protocol. The pairs Alice-Bob and Alice-Charlie perform TF-KGP separately through Eve to generate keys, while Bob and Charlie share a secret channel to Alice to exchange partial keys. In the TF-KGP, Alice-Bob and Alice-Charlie prepare signal and decoy states using a phase modulator (PM) and intensity modulator (IM), after which they send quantum signals to an untrusted party (Eve) to complete the measurement. Finally, Alice's signature is sent to Bob for authentication, and forwarded to Charlie for further verification.}
\label{fig1}
\end{figure*}

A schematic diagram of our protocol is shown in Fig. \ref{fig1}, which consists of two stages, distribution stage and messaging stage. During the distribution stage, Alice-Bob and Alice-Charlie independently implement a twin-field key generation protocol (TF-KGP) to generate correlated bit strings, where Alice-Bob and Alice-Charlie send twin-field states to the untrusted party (Eve), and Eve performs a projection measurement. The messaging stage involves sending and signing classical messages, where Alice is the sender and Bob and Charlie are the two recipients. During the distribution stage, we adopt the SNS protocol \cite{SNSTF1,SNSTF2}.

\emph{\textbf{Distribution stage:}} (1) Alice-Bob and Alice-Charlie individually generate $N$ photon pulses and code them using a phase modulator (PM) and intensity modulator (IM). During this process, each pulse is randomly chosen as the $X$ (decoy) or $Z$ (signal) window. In the $X$ window, each side randomly prepares and sends out a phase-randomized coherent state with intensity $x$, $x\in\{0,w,v\}$. In the Z window, a phase-randomized coherent state with intensity $u$ is sent with a probability $p_s$, and nothing is sent with probability $1-p_s$.

(2) Eve carries out the projection measurement on the received pulse pairs with a beam-splitter (BS) and two detectors (denoted as $D_0$, $D_1$), and publicly announces the detection results. If one of two detectors clicks, it is recorded as a successful event.

(3) Alice-Bob (Alice-Charlie) publicly announce their windows used for each pulse pair. The only successful measurement results kept are ones in which they use the same windows. Furthermore, if they both use $X$ windows, the phase and decoy states of each pulse should also be disclosed.

(4) Alice-Bob (Alice-Charlie) use the data on $Z$ windows to extract sifted keys and the data on $X$ windows to estimate parameters. In addition, they randomly sacrifice a small number of bits on the $Z$ windows for an error rate test, leaving the remainder as a signature key pool.

(5) For a future possible message $m$ ($m=0$ or $1$), Alice-Bob (Alice-Charlie) choose a length-$L$ block from the key pool to form the signature sequence $A_m^B$ and $B_m^A$ ($A_m^C$ and $C_m^A$), where $A_m^B$ and $A_m^C$ are held by Alice, and $B_m^A$ ($C_m^A$) are held by Bob (Charlie).

(6) Bob and Charlie randomly choose half of their own key bits to exchange through the Bob-Charlie secret channel. The kept half is denoted as $ B_{m,keep}^A$ ($C_{m,keep}^A$ ), and the other half as $B_{m,forward}^A$ ($C_{m,forward}^A$). Bob's and Charlie's symmetrized keys were labeled as $S_m^B=(B_{m,keep}^A,C_{m,forward}^A)$ and $S_m^C=(C_{m,keep}^A, B_{m,forward}^A)$, respectively.

\emph{\textbf{Messaging stage:}} (7) Alice sends the signature, $(m,Sig_m)$, to a recipient (such as Bob), where $Sig_m=( A_m^B, A_m^C)$.

(8) Bob compares his $S_m^B$ with $(m,Sig_m)$ and records the number of mismatches. If the mismatches are fewer than $s_a L/2$ in both key halves, Bob accepts the message and goes to the next step; otherwise, he rejects the message and aborts this round. Here, $s_a$ is the authentication threshold associated with the security level of the QDS protocol.

(9) Bob forwards $(m,Sig_m)$ to Charlie.

(10) Charlie also checks the forwarded message in the same way, but with another threshold $s_v$ ($s_v > s_a$). Charlie accepts the forwarded message if the number of mismatches in both key halves is below $s_v L/2$.

TF-KGP includes steps $(1)-(4)$ in the distribution stage. It is essentially the quantum portion of the SNS TF-QKD scheme but without error correction and privacy amplification. Detailed definitions of TF-KGP are presented in Appendix \ref{App1}, including sifted key size, error test keys, and the key pool as $n_Z$, $n_{test}$ and $n_{pool}$, respectively, and $n_Z=n_{test}+n_{pool}$. In step (6) of the distribution stage, we assume the key exchange (also called key symmetrization) between Bob and Charlie is through the Bob-Charlie secret channel, which can be realized with a TF-QKD process performed by Bob and Charlie.

\subsection{Security analysis}\label{Security}
In a QDS, although all components on $Z$ windows are used to generate keys for signature, security still depends on the single-photon components. In TF-QDS, the min-entropy resulting from single-photon components in the half of keys kept by Bob or Charlie at the presence of Eve is
\begin{align}\label{Hmin}
H_{\min }^\epsilon (\left. U^A_{m,keep} \right|E) & \geqslant  \underline{n}_{L,1}[1 - H_2(\overline{e}_{L,1})],
\end{align}
where $U\in \{B,C\}$ denotes user Bob or Charlie, and $E$ refers to the system of Eve; $\underline{n}_{L,1}$ and $\overline{e}_{L,1}$ represent the lower bound of single-photon counts and the upper bound of single-photon error rate in $U^A_{m,keep}$, respectively; $H_2(x)=-x\log_2(x)-(1-x)\log_2(1-x)$ is the binary Shannon entropy function. Eq. (\ref{Hmin}) uses a probability of $1-\epsilon$, where $\epsilon$ stands for the failure probability of the estimated parameters. In Appendix \ref{App2}, the derivations of these quantities are explained. With Eq. (\ref{Hmin}), the minimum rate $P_e$ at which Eve can introduce errors in $U^A_{m,keep}$ (length $L/2$) can be evaluated as
\begin{align}\label{Pe}
H_2(P_e) = \frac{{2\underline{n}_{L,1}}}{L}[1 - H_2(\overline{e}_{L,1})] .
\end{align}

When doing security analysis of a TF-QDS protocol, robustness, forging, and repudiation probabilities should be evaluated \cite{Amiri}. Robustness indicates the probability of the QDS aborting when Alice, Bob, and Charlie are all honest, which is caused by an error test failure. Through the error rate of test keys $E_{test}$, we can estimate the error rate in $U^A_{m,keep}$ with the Serfling inequality \cite{Serfling} using
\begin{align}\label{PET}
E_{keep}^U \leqslant E_{test}^U+\frac{2}{L}\sqrt {\frac{{\left( {\frac{L}{2} + 1} \right)\left( {\frac{L}{2} + {n_{test}}} \right)\ln \left( {\frac{1}{{{\epsilon_{PE}}}}} \right)}}{{2{n_{test}}}}},
\end{align}
except with a failure probability $\epsilon_{PE}$ and $\overline{E}_{keep}=\text{max}\{E_{keep}^B,E_{keep}^C\}$.
Considering there are failure possibilities for both processes (Alice-Bob and Alice-Charlie), robustness probability can be expressed as
\begin{align}\label{PRobust}
{\rm{P(Robust)}} \leqslant 2\epsilon_{PE}.
\end{align}

The repudiation probability characterizes Alice's signature accepted by Bob but rejected by Charlie. To repudiate, Alice must make the mismatch rate between both elements of $S_m^B$ and the signature $(m,Sig_m)$ lower than $s_a$. In addition, Alice needs the mismatch rate between either element of $S_m^C$ and the signature $(m,Sig_m)$ to be higher than $s_v$ after the key exchange. The best strategy for Alice is to control the error rate of Bob and Charlie as $E_{keep}^B=E_{keep}^C=\frac{1}{2}(s_a+s_v)$ \cite{Amiri}, in which case the repudiation probability is bounded by
\begin{align}\label{PRep}
{\rm{P}}({\rm{Repudiation}}) \leqslant 2{e^{ - \frac{1}{4}{{\left( {{s_v} - {s_a}} \right)}^2}L}},
\end{align}
where ${s_a} = {\overline{E}_{keep}} + {{({P_e} - {\overline{E}_{keep}})} \mathord{\left/{\vphantom {{({P_e} - {\overline{E}_{keep}})} 3}} \right.\kern-\nulldelimiterspace} 3}$, and ${s_v} = { \overline{E}_{keep}} + 2{{({P_e} - {\overline{E}_{keep}})} \mathord{\left/
 {\vphantom {{({P_e} - {\overline{E}_{keep}})} 3}} \right.\kern-\nulldelimiterspace} 3}$

The forging indicates that the signature is not signed by Alice but would be accepted by Bob and Charlie. For simplicity, we assume Bob is a forger. In order to forge, Bob must keep the mismatch rate between his declaration $(m,Sigm)$ and Charlie's keys ($S_m^C=(C_{m,keep}^A,B_{m,forward}^A)$) being lower than a given value $s_v$. Considering half of Charlie's string $B_{m,forward}^A$ is forwarded by Bob, Bob needs only to guess the left half $C_{m,keep}^A$.
The forging probability includes all the process guessing $C^A_{m,keep}$, which is written as
\begin{align}\label{PForge}
{\rm{P(Forge)}} \leqslant g + \epsilon_{F} + \epsilon_{PE} + \epsilon_{\underline{n}_{L,1}} + \epsilon_{\overline{e}_{L,1}},
\end{align}
where $g$ and $\epsilon_F$ are associated with the probability that Bob finds a signature with an error rate smaller than $s_v$, defined by
\begin{align}\label{epsilon}
{\epsilon_F}: = \frac{1}{g}\left( {{2^{ - \frac{L}{2}\left\{ {\frac{{2\underline{n}_{L,1}}}{L}[1 - H_2(\overline{e}_{L,1})] - {H_2}({s_v})} \right\}}} + \epsilon} \right).
\end{align}
$\epsilon_{PE}$, $\epsilon_{\underline{n}_{L,1}}$ and $\epsilon_{\overline{e}_{L,1}}$ are the error probabilities related to the estimation of $\overline{E}_{keep}$, $\underline{n}_{L,1}$, and $\overline{e}_{L,1}$, respectively.
To define $\varepsilon$ as the security level of the system, according to \cite{Collins2017} it requires
\begin{align}\label{varepsilon}
 \text{max}\{ {\rm{P(Robust)}},{\rm{P}}({\rm{Repudiation}}), {\rm{P(Forge)}}\} \leqslant \varepsilon.
\end{align}

We now present a simple model to evaluate the performance of a QDS protocol with the desired security level $\varepsilon$. We assume that for each run, the block of photon pulse pairs ($N$) to perform key distribution is known, and set $n_{pool}$ as the corresponding number of keys that can be used for signing. Subsequently we can calculate how many keys are needed to sign a half-bit signature ($L$), and how many bits should be signed ($n_{bits}$) with $n_{pool}$ keys. Finally, the signed bits and signature rate ($bit/pulse$) can be written as
\begin{align}\label{BitsRate}
n_{bits} = \frac{{{n_{pool}}}}{{2L}}, \\
R = \frac{{{n_{pool}}}}{{2L}}\cdot\frac{1}{N},
\end{align}
respectively.

\section{Numerical simulations}
\begin{figure}[h!]
\centering
\includegraphics[width=10cm]{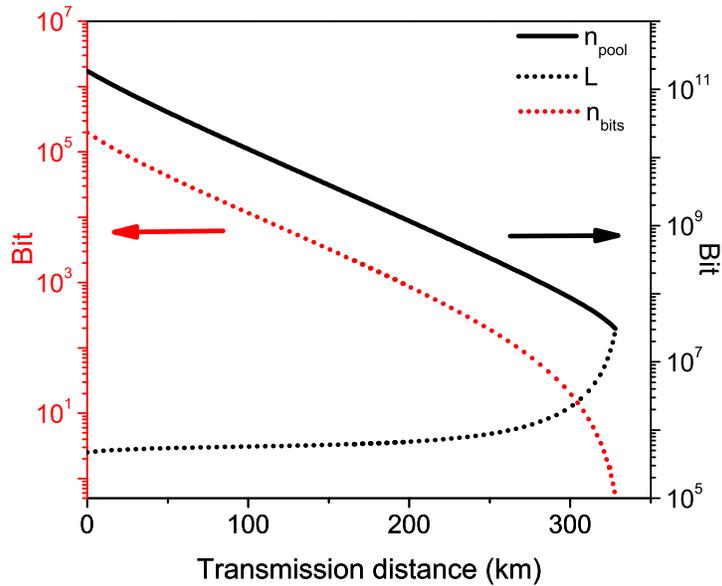}
\caption{The size of key pool ($n_{pool}$), the length for signing message $m$ ($L$), and the number of signed bits ($n_{bits}$) versus the total transmission distance. $n_{pool}$ and $L$ correspond to the right axis while $n_{bits}$ is indicated on the left axis. The data size is $N=10^{13}$ and the security level is $\varepsilon=10^{-5}$ at all transmission distances.}
\label{fig2}
\end{figure}

\begin{figure}[htbp]
\centering
\includegraphics[width=12cm]{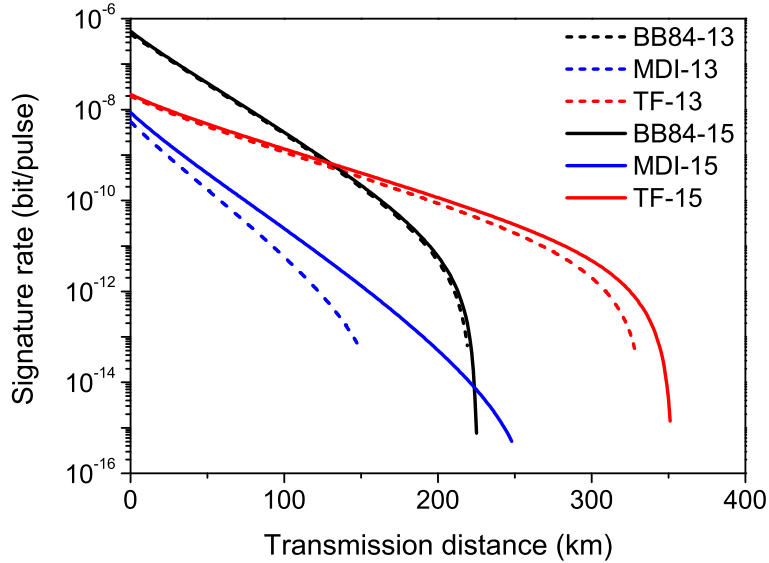}
\caption{Signature rates of BB84-QDS \cite{Amiri}, MDI-QDS \cite{MDIQDS}, and TF-QDS with a security level of $\varepsilon=10^{-5}$. The dashed lines represent results at data size $N=10^{13}$, and the solid lines at data size $N=10^{15}$. }
\label{fig3}
\end{figure}

\begin{figure}[htbp]
\centering
\includegraphics[width=12cm]{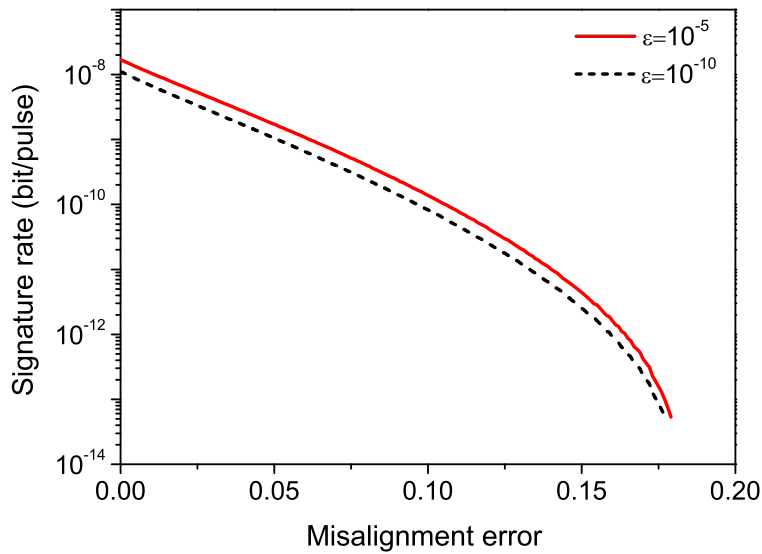}
\caption{Signature rates of TF-QDS versus optical misalignment errors with $\varepsilon=10^{-5}$ or $\varepsilon=10^{-10}$ at 50 km. Here, $N=10^{13}$. }
\label{fig4}
\end{figure}

In this section, we describe numerical simulations for our proposed TF-QDS protocol, with results shown in Figs. \ref{fig2}-\ref{fig4}. In our simulations, we consider statistical fluctuations using Hoeffdings inequality \cite{Hoeffding} as in Ref. \cite{Amiri}, and the basic system parameters are listed in Table \ref{tab1}. In addition, we set the number of phase slices to $M=16$ \cite{TFQKD} during the TF-KGP processes, where the phase slice is the coding phase difference interval post-selected by Alice-Bob or Alice-Charlie. In addition, we perform full parameter optimization on our TF-QDS, including the value of each light intensity ($w, v, u$), probability of choosing signal window ($p_Z$), probability of sending out the phase-randomized coherent state $u$ ($p_s$), and probabilities of choosing different decoy intensities ($p_{w}$ and $p_{v}$).
\begin{table}[htbp]
  \caption{The basic system parameters used in our numerical simulations. $\alpha$: the loss coefficient of fiber at telecommunication wavelength; $\eta_d$ and $P_{dc}$: detection efficiency and the dark count rate of detectors, respectively; $e_d$: optical misalignment error; $r_{ET}$: the ratio of keys used for the error test; $\epsilon_{PE}$ and $\epsilon_{SF}$: failure probability of the error test and statistical fluctuation, respectively; $g$: Bob's probability of making $s_vL/2$ errors.}
\renewcommand{\arraystretch}{1.3}
\begin{tabularx}{\linewidth}{XXXXXXXX}  \hline \hline
$\alpha$ & $\eta_d$ &  $P_{dc}$ & $e_d$ &  $r_{ET}$ &  $\epsilon_{PE}$  & $\epsilon_{SF}$ &  $g$  \\ \hline
$0.2dB/km$ & 50\% &  $10^{-7}$ & 0.03 & 5.5\% & $10^{-12}$  &  $10^{-12}$ &  $10^{-12}$  \\ \hline\hline
  \end{tabularx}
 \label{tab1}
\end{table}

In Fig. \ref{fig2}, we illustrate the variation of $n_{pool}$, $L$, and $n_{bits}$ with varying transmission distances, given a data size of $N=10^{13}$ and a security level of $\varepsilon=10^{-5}$, where $n_{bits}$ is shown with the left axis, and $n_{pool}$ and $L$ are indicated by the right axis. The value of $L$ increases rapidly and the value of $n_{bits}$ drops quickly as transmission distance increases, especially after 300 km. These observations can be attributed to the finite-size effect, which is more sensitive at longer transmission distances.

The signature rate of TF-QDS is plotted in Fig. \ref{fig3}, and compared to two typical QDS protocols, BB84-QDS \cite{Amiri} and MDI-QDS \cite{MDIQDS}. We set the security level as $\varepsilon=10^{-5}$ and the data size as $N=10^{13}$ or $N=10^{15}$ \cite {MDI404}. For fair comparisons, we also perform full parameter optimization on BB84-QDS and MDI-QDS. We can see from Fig. \ref{fig3} that among the three protocols, our TF-QDS protocol exhibits the best performance at longer transmission distances. For example, our TF-QDS can sign signatures at 300 km while the other two protocols stop signing at 230 km and 250 km, respectively. At shorter transmission distances, BB84-QDS exhibits the highest signature rate. However, it possesses the lowest security level among the three protocols. Therefore, when taking security level into account, our protocol exhibits the best performance in terms of both transmission distance and signature rate.

We also investigate the robustness of our TF-QDS protocol by plotting signature rate variations with changes in the misalignment of the optical system in Fig. \ref{fig4}. Here, the transmission distance is set at 50 km and the security level as $\varepsilon=10^{-5}$ or $\varepsilon=10^{-10}$.
Fig. \ref{fig4} shows that the signature rate decreases with increasing misalignment error. The maximum tolerable misalignment error is 18\%, which is much larger than values in the BB84 and MDI protocols, and well within current experimental values \cite{Minder,WangS,LiuY,ZhongX}. Moreover, by setting a higher security level, a lower signature rate can be obtained. Therefore, a reasonable security level should be chosen in practical applications of the TF-QDS. In addition, due to the phase sensitivity of TF-KGP, we can use machine learning to achieve phase-modulation stabilization \cite{LiuJY}, enhancing the practical performance of the TF-QDS system.

\section{Conclusions}
In this paper, we develop a TF-QDS protocol, which can possess the highest security level among all existing QDS protocols, but also exhibit outstanding performance in terms of both signature rates and secure transmission distances. For example, the TF-QDS protocol can achieve $>$ 100 km longer secure transmission distance than either BB84-QDS or MDI-QDS under the same experimental conditions, and exhibits a higher signature rate than MDI protocol by several orders of magnitude after 200 km. Therefore, our work represents another step towards practical implementation of QDS.

To be noted, this is the first TF-QDS protocol, by adopting Wang \emph{et al.}'s SNS scheme \cite{SNSTF1,SNSTF2}. In principle, other types of TF schemes \cite{CuiC,Curty,MCSTF,AsyTF} and security analysis methods \cite{Maeda,Lorenzo} could also be implemented in QDS, and might show even more interesting characteristics. This will be carried out in our future research work. As for the limitations of the present QDS work, similar to existing TF-QKD protocols, it might pose high challenges for wide applications in the field, e.g., it needs high speed and accurate multi-party synchronization, phase-locking and stabilization techniques. Anyway, with the rapid development of modern technology, all these challenges will be readily solved. Therefore, our work represents another step towards practical implementation of QDS.

\section*{ACKNOWLEDGMENTS}
We also acknowledge financial support from the National Key Research and Development Program of China (Grants No. 2018YFA0306400, No. 2017YFA0304100); National Natural Science Foundation of China (NSFC) (Grants No. 11774180, No. 61590932, No. 61705110, No. 11847215); China Postdoctoral Science Foundation (Grant No. 2018M642281).

\section*{COMPETING INTERESTS}
The authors declare that there are no competing interests.

\appendix
\section{Some detailed notes on the TF-KGP}\label{App1}
In this section, we provide detailed notes on the TF-KGP. We start by analyzing the TF-KGP procedure between Alice and Bob, which is the SNS TF-QKD presented in \cite{SNSTF1,SNSTF2} without error correction and privacy amplification.

The phase-randomized coherent state prepared by Alice and Bob, respectively, can be expressed as
\begin{align}\label{Coherent}
\left| {\sqrt x_A {e^{i\theta_A }}} \right\rangle  = \sum\nolimits_{n = 0}^\infty  {\frac{{{e^{ - x_A/2}}{{(\sqrt x_A {e^{i\theta_A }})}^n}}}{{\sqrt n !}}\left| n \right\rangle } , \quad
\left| {\sqrt x_B {e^{i\theta_B }}} \right\rangle  = \sum\nolimits_{n = 0}^\infty  {\frac{{{e^{ - x_B/2}}{{(\sqrt x_B {e^{i\theta_B }})}^n}}}{{\sqrt n !}}\left| n \right\rangle } ,
\end{align}
where $x_A$ ($x_B$) and $\theta_A$ ($\theta_B$) represent the intensity and phase of coherent state randomly chosen by Alice (Bob), respectively. Here, $x_A, x_B \in \{0, w, v, u\}$ and $\theta_A, \theta_B$ are random in $[0,2\pi)$. Alice (Bob) randomly chooses a vacuum state, decoy states ($w$, $v$), and $Z$ windows with probabilities $p_0, p_w, p_v, p_Z$, respectively, where $p_0+p_w+p_v+p_Z=1$. When a $Z$ window is chosen, Alice (Bob) sends a signal state $u$ with probability $p_s$, and sends nothing with $1-p_s$.

When receiving the pulses from Alice and Bob, Charlie performs measurements and announces the results. During the measurement process, if only one detector clicks, Charlie announces a successful event, recorded as a one-detector heralded event, and announces which detector ($D_0$ or $D_1$) clicks. When the measurement process is complete and results have been announced, Alice and Bob publicly disclose which window was used for each pulse pair. Only the one-detector heralded events for which they both use $X$ or $Z$ windows are kept. When they both use $X$ windows, the phase and decoy state intensity should also be disclosed. However, when they both use $Z$ windows, the phase and SNS operation should be never disclosed. Furthermore, we need to post-select the effective events on $X$ windows; it is deemed an effective event if it is a one-detector heralded event where Alice and Bob both use $X$ windows, the two coherent states of Alice and Bob have the same intensity, and their phases satisfy the following post-selection criterion
\begin{align}\label{PhaseSlice}
\left|\theta_{A}-\theta_{B}-\psi_{\mathrm{AB}}-k\pi\right| \leq \frac{\Delta}{2}.
\end{align}
In Eq. (\ref{PhaseSlice}), $\psi_{\mathrm{AB}}$ is the difference of global phases between Alice-Eve's link and Bob-Eve's link, which results in the optical misalignment error ($e_d$); $k=0,1$ corresponds to in-phase or anti-phase of ${\theta _A}$ and ${\theta _B}$; $\Delta=\frac{{2\pi }}{M}$ represents the size of each slice, and $M$ refers to the total number of phase slices pre-chosen by Alice and Bob. The effective events on $X$ windows are the results of the single-photon interference and a subset of one-detector heralded events on $X$ windows.

For the one-detector heralded events on $Z$ windows, Alice (Bob) denotes it as bit 0 if she (he) sends a vacuum (phased-randomized weak coherent) state and as bit 1 if she (he) sends a phased-randomized weak coherent (vacuum) state. For the effective events on $X$ windows, a right click is the $D_0$ ($D_1$) detector clicking when $k=0$ ($k=1$), and a wrong click is the $D_1$ ($D_0$) detector clicking when $k=0$ ($k=1$).
The data on $Z$ windows are defined as the key bits distilled by the one-detector heralded events on $Z$ windows, while the data on $X$ windows are defined as the one-detector heralded and effective events on $X$ windows. The data on $Z$ windows are used for the error test and signature, and finally Alice and Bob form an $n_Z$-length key string $Z_s$ and $Z^\prime_s$, respectively. The data on $X$ windows are used to estimate single-photon contributions, i.e. the counts and error rates of the single-photon components ($\underline{n}_{L,1}, \overline{e}_{L,1}$) on $Z$ windows.

\section{Finite-size estimations of parameters}\label{App2}
In this section, we estimate $\underline{n}_{L,1}$ and $\overline{e}_{L,1}$ in finite size. The procedure can be decomposed into three steps.

Firstly, we estimate the lower bound of single-photon counts and upper bound of single-photon error counts on $X$ windows ($\underline{n}_{X,1}$ and $\overline{m}_{X,1}$, respectively) with the observed values taking statistical fluctuations into account. From the data on $X$ windows, we know the counts of one-detector heralded events with various intensity combinations ($n_{ab}$, $a,b \in \{0,w,v\}$), and the counts of error clicks in effective events $m_{aa}$. With these observed values, we obtain
\begin{align}\label{n1X}
{n_{X,1}} \geqslant  \underline{n}_{X,1} = \frac{{{\tau _{X,1}}}}{{2wv(v - w)}}\left[ {\frac{{{v^2}{e^w}({n^-_{0w}} + {n^-_{w0}})}}{{{P_{0w}}}} - \frac{{{w^2}{e^v}({n^+_{0v}} + {n^+_{v0}})}}{{{P_{0v}}}} - \frac{{2({v^2} - {w^2}){n^+_{00}}}}{{{P_{00}}}}} \right],
\end{align}
\begin{align}\label{m1X}
{m_{X,1}} \leqslant {\overline{m}_{X,1}} = \frac{{{\tau _{X,1}}}}{{v - w}}\left[ {\frac{{{e^v}{m^+_{vv}}}}{{P_{vv}^\Delta }} - \frac{{{e^v}{m^-_{ww}}}}{{P_{ww}^\Delta }}} \right],
\end{align}
and the corresponding single-photon error rate on $X$ windows is $\overline{e}_{X,1}=\overline{m}_{X,1}/\underline{n}_{X,1}$. In Eqs. (\ref{n1X}) and (\ref{m1X}), $\tau _{X,1}$ is the probability of single-photon components with all intensity combinations on $X$ windows, which is ${\tau _{X,1}} = \sum\nolimits_{a,b} {{P_{ab}}(a + b){e^{ - a - b}}}$. $P_{ab}$ is the probability of intensity combination $ab$, and $P_{aa}^\Delta$ is the probability of effective events occurring with intensity combination $aa$, given by $P_{ab}=p_ap_b$ and $P_{aa}^\Delta = 2p_a^2\frac{\Delta }{{2\pi }}$.
The $x^-$ and $x^+$ in Eqs. (\ref{n1X}) and (\ref{m1X}) are the observed values when considering the statistical fluctuations by the Hoeffding inequalities \cite{Hoeffding}
\begin{align}\label{HoeffdingEq}
{\tilde x} \geqslant x^- := x -  \delta (x,\epsilon_{SF}) ,\quad {\tilde x} \leqslant x^+ := x +  \delta (x,\epsilon_{SF}),
\end{align}
with failure probability $\epsilon_{SF}$, where
\begin{align}\label{HoeffdingDelta}
\delta (x,\epsilon_{SF}) = \sqrt {\frac{{x\ln (1/{\epsilon_{SF}})}}{2}}.
\end{align}

Secondly, since the single-photon signals on $X$ and $Z$ windows are independent, we use $\underline{n}_{X,1}$ and $\overline{m}_{X,1}$ to estimate the corresponding quantities on $Z$ windows ($n_{Z,1}$ and $m_{Z,1}$) using the Serfling inequality \cite{Serfling}. The population for single-photon preparations on $Z$ windows is lower bounded by
\begin{align}\label{Nz1}
{\underline{N}_{Z,1}} = 2{p_s}(1 - {p_s})u{e^{ - u}}{N_Z} - \delta ({N_Z},{\epsilon_{SF}}),
\end{align}
with confidence $1-\epsilon_{SF}$, where $N_Z=p_Z^2N$ represents the runs of Alice and Bob both choosing the $Z$ windows, and $\delta(x,y)$ is the fluctuation in Hoeffding's inequality \cite{Hoeffding}. Similarly, the population for single-photon preparations on $X$ windows is upper bounded by
\begin{align}\label{Nx1}
{\overline{N}_{X,1}} = \sum\nolimits_{a,b} {\left[ {(a + b){e^{ - a - b}}{N_{ab}} + \delta ({N_{ab}},{\epsilon_{SF}})} \right]},
\end{align}
with confidence $1-9\epsilon_{SF}$, where $N_{ab}=P_{ab}N$ represents Alice and Bob choosing an intensity combination $ab$ on the $X$ windows. Subsequently, we can interpret the single-photon contributions on $X$ or $Z$ windows ($n_{X,1}$, $m_{X,1}$, $n_{Z,1}$,$m_{Z,1}$) in the whole population of the single-photon preparations as an operation of sampling without replacement. The Serfling inequality tells us that
\begin{align}\label{SerflingEq1}
{n_{Z,1}} \geqslant {\underline{n}_{Z,1}} = {\underline{n}_{X,1}}\frac{{{\underline{N}_{Z,1}}}}{{{\overline{N}_{X,1}}}} - \Upsilon ({\underline{N}_{Z,1}},{\overline{N}_{X,1}},{\epsilon_{SF}}),\\
{m_{Z,1}} \leqslant {\overline{m}_{Z,1}} = {\overline{m}_{X,1}}\frac{{{\underline{n}_{Z,1}}}}{{{\underline{n}_{X,1}}}} + \Upsilon ({\underline{n}_{Z,1}},{\underline{n}_{X,1}},{\epsilon_{SF}}),
\end{align}
where confidence is $1-\epsilon_{SF}$, and $\Upsilon(x,y,z)= \sqrt {(x + 1)(x + y)\ln ({z^{ - 1}})/(2y)}$. The corresponding single-photon error rate on $Z$ windows is
\begin{align}\label{eZ1}
\overline{e}_{Z,1}=\frac{{{\overline{m}_{Z,1}}}}{{{\underline{n}_{Z,1}}}}.
\end{align}

Thirdly, we can use $\underline{n}_{Z,1}$ and $\overline{e}_{Z,1}$ to estimate $n_{L,1}$ and $e_{L,1}$ in $U^A_{m,keep}$ through the Serfling inequality with
\begin{align}\label{nL1eL1}
{n_{L,1}} &\geqslant {\underline{n}_{L,1}} = {\underline{n}_{Z,1}}\frac{{{L}}}{{{2n_{Z}}}} - \Lambda ({n_{Z}},{\frac{L}{2}},{\epsilon_{SF}}),\\
{e_{L,1}} &\leqslant {\overline{e}_{L,1}} = {\overline{e}_{Z,1}} + {\frac{1}{\underline{n}_{L,1}}}\Lambda ({\underline{n}_{Z,1}},{\underline{n}_{L,1}},{\epsilon_{SF}}),
\end{align}
where $\Lambda(x,y,z)=\sqrt {(x - y + 1)y\ln ({z^{ - 1}})/(2x)}$. Finally, we obtain $\underline{n}_{L,1}$ and $\overline{e}_{L,1}$.

In addition, we simulate the experimental observed values with the linear model presented in \cite{SNSTF2} and assume symmetric case in the TF-KGP. If the total transmittance of the experiment setups is $\eta=\eta_d 10^{-\frac{\alpha }{{20}}}$, then we have
\begin{align*}
n_{00}=&2 P_{dc}\left(1-P_{dc}\right) N_{00},\\
n_{0w}=&n_{w0}=2\left[\left(1-P_{dc}\right) e^{\eta w / 2}-\left(1-P_{dc}\right)^{2} e^{-\eta w}\right] N_{0w},\\
n_{0v}=&n_{v0}=2\left[\left(1-P_{dc}\right) e^{\eta w / 2}-\left(1-P_{dc}\right)^{2} e^{-\eta w}\right] N_{0v},\\
m_{w w}=&e_{d}\left[\left(1-P_{d c}\right)\frac{1}{\Delta} \int_{-\frac{\Delta}{2}}^{\frac{\Delta}{2}} e^{-2 \eta w \sin ^{2} \frac{\theta_{AB}}{2}} d \theta_{AB}-\left(1-P_{d c}\right)^{2} e^{-2 \eta w}\right]P_{ww}^\Delta N \\  &+\left(1-e_{d}\right)\left[\left(1-P_{d c}\right)\frac{1}{\Delta} \int_{-\frac{\Delta}{2}}^{\frac{\Delta}{2}} e^{-2 \eta w \cos ^{2} \frac{\theta_{AB}}{2}} d \theta_{A B}-\left(1-P_{d c}\right)^{2} e^{-2 \eta w}\right] P_{ww}^\Delta N, \\
m_{vv}=&e_{d}\left[\left(1-P_{d c}\right)\frac{1}{\Delta} \int_{-\frac{\Delta}{2}}^{\frac{\Delta}{2}} e^{-2 \eta v \sin ^{2} \frac{\theta_{A B}}{2}} d \theta_{AB}-\left(1-P_{d c}\right)^{2} e^{-2 \eta v}\right]P_{vv}^\Delta N \\  &+\left(1-e_{d}\right)\left[\left(1-P_{d c}\right)\frac{1}{\Delta} \int_{-\frac{\Delta}{2}}^{\frac{\Delta}{2}} e^{-2 \eta v \cos ^{2} \frac{\theta_{AB}}{2}} d \theta_{AB}-\left(1-P_{d c}\right)^{2} e^{-2 \eta v}\right] P_{vv}^\Delta N, \\
\end{align*}
and
\begin{align*}
{n_Z} = &2{(1 - {p_s})^2}{P_{dc}}(1 - {P_{dc}}){N_Z} + 4{p_s}(1 - {p_s})\left[ {(1 - {P_{dc}}){e^{ - \eta u/2}} - {{(1 - {P_{dc}})}^2}{e^{ - \eta u}}} \right]{N_Z} \\
&+ 2p_s^2\left[ {(1 - {P_{dc}}){e^{ - \eta u}}\frac{1}{{2\pi }}\int_0^{2\pi } {{e^{\eta u\cos {\theta _{AB}}}}d{\theta_{AB}}}  - {{(1 - {P_{dc}})}^2}{e^{ - 2\eta u}}} \right]{N_Z},
\end{align*}
where $\theta_{AB} = \theta_{A} - \theta_{B}$.

\end{document}